# A Gold Standard for Emotion Annotation in Stack Overflow

Nicole Novielli, Fabio Calefato, Filippo Lanubile
University of Bari Aldo Moro, Italy
{nicole.novielli, fabio.calefato, filippo.lanubile}@uniba.it

## ABSTRACT

Software developers experience and share a wide range of emotions throughout a rich ecosystem of communication channels. A recent trend that has emerged in empirical software engineering studies is leveraging sentiment analysis of developers' communication traces. We release a dataset of 4,800 questions, answers, and comments from Stack Overflow, manually annotated for emotions. Our dataset contributes to the building of a shared corpus of annotated resources to support research on emotion awareness in software development.

## CCS CONCEPTS

• **Information systems** → **Retrieval tasks and goals**; *Sentiment Analysis*; • **Software creation and management** → Collaboration in software development

## KEYWORDS

Emotion Mining, Sentiment Analysis, Communication Channels, Stack Overflow, Social Software Engineering.

**ACM Reference format:**

N. Novielli, F. Calefato, F. Lanubile. 2018. A Gold Standard For Emotion Annotation in Stack Overflow. In *Proceedings of MSR 2018*, 4 pages. DOI: …[1]

## 1 INTRODUCTION

Recent research has provided evidence that software developers experience a wide range of emotions and express them throughout a rich ecosystem of communication channels [4][13][15]. As such, researchers have started to study the role of affective states in software engineering, by applying sentiment analysis to crowd-generated content within social software engineering tools [5][11][16][18]. However, identifying the positive, negative, or neutral polarity of a text, is only one of the possible dimensions of affect analysis. Thus, recent research advocates in favor of the emergence of sentiment analysis tools monitoring communication between the developers at a more fine-grained level of analysis, by detecting specific affective states, such as emotions or attitudes (e.g., joy, anger) [13], as well as their target (e.g., a tool, a teammate) [5], so as to enable the translation of emotion mining into actionable insights.

To support empirical research in this direction, we developed a gold standard dataset [2] collected from Stack Overflow and manually annotated with emotion labels. Since mining emotions from text requires an appropriate model to operationalize sentiment [13], we defined our annotation guidelines based on the framework by Shaver et al. [17], which has been previously adopted in empirical studies on emotion awareness in software development [2][3][11][15]. The framework defines a tree-structured hierarchical classification of emotions, where each level refines the granularity of the previous one, thus providing more indication of its nature. At the top level, the framework includes six basic emotions, namely love, joy, anger, sadness, fear, and surprise.

Our Stack Overflow dataset complements the effort made by Ortu et al. [15], towards the construction of a gold standard dataset to support the study of emotions in software engineering. The envisioned users of our dataset are researchers interested in investigating the role of emotions in software development, [5][11][14][16][18].

The remainder of this paper is organized as follows. In Section 2 we describe how the dataset has been created and validated, including the methodology followed to annotate the gold standard. In Section 3 we present some opportunities that the dataset offers to researchers. In Section 4 we discuss the threats to validity of the annotation study. Finally, Section 5 and Section 6 include, respectively, the related work and conclusions.

## 2 DATASET

The dataset includes 4,800 posts from Stack Overflow in the form of questions, answers, and comments. For all posts in the dataset, we distribute both the set of individual annotations provided by the raters (i.e., the indication of presence or absence for each emotion) and the gold label obtained by applying majority voting. Of 4,800 posts, 1,959 received at least an emotion label while 2,841 were marked as neutral due to the absence of an affective label. The distribution of emotion labels in our dataset is reported in Table 1 while examples of annotations are shown in Table 2.

---

[1]

[2] Gold standard and guidelines available at: https://github.com/collab-uniba/EmotionDatasetMSR18

**Table 1. Emotion label distribution.**

| Posts | Texts conveying the emotion | | | | | | N |
|---|---|---|---|---|---|---|---|
| | *Love* | *Joy* | *Surprise* | *Anger* | *Sadness* | *Fear* | |
| # | 1,220 | 491 | 45 | 882 | 230 | 106 | 4,800 |
| % | 25% | 10% | 1% | 18% | 5% | 2% | |

**Table 2. Examples of Annotated Posts.**

| Input text | Annotation | |
|---|---|---|
| | **Basic Emotion** | **Rationale for annotation** *(second and/or third level emotion found)* |
| *"Thanks for your input! You're, like, awesome!"* | Love | Liking (third level), Affection (second level) indicating gratitude. |
| *"I'm happy with the approach, the code looks good"* | Joy | Happiness, Satisfaction (third), Cheerfulness (second) |
| *"Absolutely terrible API design"* | Anger | Dislike (third), Rage (second) |

Please note that the percentages do not sum up to 100% because multiple labeling of posts is allowed. In particular, 133 posts (3%) received two emotion labels, with the most frequent couple of labels being love and joy (70 items).

The annotation sample was extracted from the official Stack Overflow data dump from July 2008 to September 2015. A previous emotion annotation study showed how the large proportion of text contributed by developers is neutral, i.e. does not contain any trace of emotion [11]. Therefore, we built the dataset for the annotation by performing opportunistic sampling of posts based on both the presence of affectively-loaded lexicon and their type. The purpose of opportunistic sampling is twofold: on one hand, we want to avoid wasting the time of raters by asking them to manually label mainly neutral posts; on the other hand, we aim at obtaining a dataset in which positive and negative emotions, as well as absence of them, are equally represented in the data. Thus, we used SentiStrength [19] to assess the presence/absence of affective lexicon in a post, as done in previous research [3]. We computed the positive and negative sentiment scores for the text of all the four types of posts extracted from the StackOverflow dump. Then, we randomly selected the same number of items based on the type of post (i.e., question, answer, or comment) and its sentiment scores (i.e., positive, negative, or neutral overall polarity).

Our sample for annotation contains 4,800 items overall, equally distributed with respect to the types of posts and polarity, i.e. one-third of posts scored as positive by SentiStrength, one-third as negative, and one-third as neutral. To improve their readability, we pre-processed all the posts and discarded all those elements that are out of the scope of the sentiment annotation task, e.g., code snippets, URLs, and HTML tags. We consider as a unit of analysis the Stack Overflow post, which includes questions, answers, and comments provided by community members.

**Table 3. Interrater agreement for emotion annotation.**

| | Love | Joy | Surprise | Anger | Sadness | Fear |
|---|---|---|---|---|---|---|
| **Obs. Agreement** | .87 | .86 | .98 | .88 | .93 | .96 |
| **Fleiss' Kappa** | .66 | .40 | .30 | .62 | .45 | .39 |

The dataset was annotated by twelve volunteers, recruited among graduate CS students at our university. The coders were requested to indicate the presence/absence of each emotion from the Shaver et al.'s framework. Each post was annotated by three raters and disagreements were resolved the by applying a majority voting criterion. Coders were trained in a joint 2-hour session by the first author. After explaining and discussing the coding guidelines, the training was completed with a pilot subset of 100 items, to be annotated individually at home. The twelve participants were organized into four groups of three coders each. Therefore, the pilot study was performed on 400 posts overall and each item in the dataset was assigned to three coders. A week later, the annotations were discussed in a 2-hour plenary meeting with the experimenter, to resolve the disagreements and disambiguate the unclear parts of the guidelines. After all the disagreements were solved, the pilot annotation became the first building block of the gold standard.

Once the training was completed, we assigned a new set of 500 posts to each coder. Overall, 2,000 new items were annotated in this second step. Again, each item was annotated by three coders who individually performed this new annotation task. The deadline for returning the annotation was set in three weeks. We then assigned the final set of 600 posts to the coders. Overall, 2,400 additional new items were annotated in this final step. The observed agreement values (i.e., the percentage of cases for which raters provided the same annotation) for each emotion label range from .86 for *joy* to .98 for *surprise* (see Table 3), thus demonstrating the reliability of our gold standard. For the sake of completeness and enable comparison with previous work [11], Fleiss' kappa (κ) is also reported. In spite of the almost perfect observed agreement, κ ranges from moderate to substantial agreement due to the skewed distribution of labels, with lower values observed for less frequent emotions (joy, surprise, sadness, and fear). This is an effect of the correction of the observed agreement operated by κ with respect to the chance agreement, which is higher for highly unbalanced data. Still, the values for κ are comparable to those observed by previous annotation performed by Ortu et al. using the same theoretical framework for emotions (see Figure 1).

## 3   RESEARCH OPPORTUNITIES

Off-the-shelf sentiment analysis tools, which were trained on non-technical domains, have been demonstrated to produce unreliable results in software engineering [7]. The main envisaged use for this dataset is for training and validation of emotion mining tools specifically optimized for software engineering, thus complementing the effort by Ortu et al. [15] towards building shared corpora for sentiment analysis in software engineering.

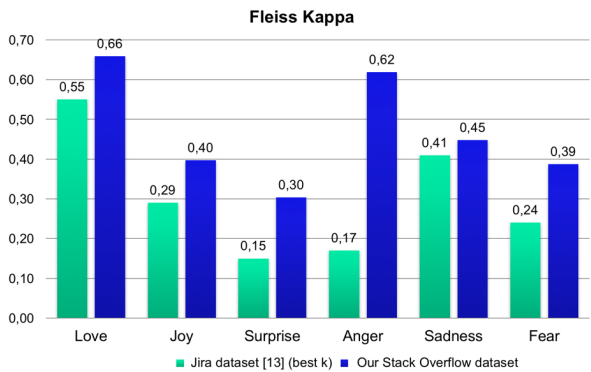

**Figure 1 - Comparison with k for the Jira dataset [11]**

In particular, our dataset has been already employed to train and validate *EmoTxt*, an open source toolkit for mining emotions from communication exchanges, structured as a suite of six binary classifiers, each predicting the presence/absence of a specific emotion in the input text. Furthermore, the mapping between gold labels for emotions and their positive, negative, and neutral polarity has been used for training a *Senti4SD* [2], which is specifically optimized for software development. Both tools are part of our *Emotion Mining Toolkit* (EMTk)[3].

Moreover, our dataset could be used for developing lexical resources for unsupervised, lexicon-based approaches such as SentiStrengthSE [6]. For example, Mantyla et al. [10] defined and implemented an approach for bootstrapping a lexicon for identifying emotional arousal, i.e., the level of emotional activation, for the software engineering domain. A similar approach could be implemented to extract emotion dictionaries from our data.

Another potential use of our gold standard is for investigating the role of emotions in collaborative knowledge building. Recent studies leveraged sentiment analysis of Stack Overflow posts to investigate the impact of emotions on the success of Stack Overflow questions [4], to summarize developers' opinion about API [20], and provide recommendations accordingly [8].

Finally, by sharing our guidelines for annotation we want to ease the execution of replications as well as new studies on emotion awareness in software engineering.

## 4 THREATS TO VALIDITY

Among the available information sources belonging to the rich social programmer ecosystem, our dataset includes data from a technical Q&A site such as Stack Overflow. We are aware that our methodology may produce different results when applied to other data sources, such as comments on social coding sites or issue tracking systems. However, Stack Overflow is so popular among software developers (currently used by more than 8 million software developers[4]) to be reasonably confident that the dataset is representative of developers' communication style.

We built our gold standard on emotion polarity through manual annotation. Emotion annotation is a subjective process since affect triggering and perception can be influenced by personality traits and personal dispositions (Scherer et al. 2004). To mitigate this threat, we provide clear guidelines grounded on a theoretical framework for emotion identification. Furthermore, final gold labels were assigned using majority agreement among three coders. The observed agreement confirms a good reliability of the gold standard.

Finally, the sample set for the emotion annotation was built through opportunistic sampling using SentiStrength to have one-third of posts scored as conveying a positive emotion, one-third conveying negative emotions, and one-third as neutral (i.e., the absence of emotions). As such, we might have filtered out text items conveying interesting emotional content using a lexicon that is not recognized by SentiStrength as emotional.

## 5 RELATED WORK

The closest dataset for emotion annotation currently available for research is the Jira dataset of developers' comments, released by Ortu et al. [15]. It has been developed by adopting the same theoretical framework by Shaver that we used for annotating our Stack Overflow corpus. Their gold standard was used by its authors to train an emotion classifier [14] and by other researchers as a gold standard for benchmarking off-the-shelf sentiment analysis tools [7].

As far as emotion polarity detection is concerned, researchers worked towards overcoming the limitations posed by off-the-shelf sentiment analysis tools [7], such as SentiStrength [19], when trained outside the software engineering domain. Islam and Zibran [6] have developed Sentistrength-SE, a software engineering-specific version of SentiStrength [19] incorporating *ad hoc* heuristics and adjusted sentiment scores of words in its lexicon. Optimization is based on the performance observed on a small dataset of 400 developers' comments in Jira [15]. Ahmed et al. have released SentiCR [1], a sentiment analysis tool trained on a manually annotated gold standard of 2,000 code review comments. A Stack Overflow dataset was collected in the scope of a broader study aiming at developing a recommender for software libraries, which leverages sentiment analysis for mining crowdsourced opinions [8].

Other dimensions of affect have also been studied. It is the case of Mäntylä et al. [9] who proposed an approach to score the developers' emotions in Jira along with valence, arousal, and dominance, for the early detection of burnout in open source projects. Gachechiladze et al. [5] reported about a preliminary investigation of supervised approaches to detect the target of anger in developers' communication in Jira.

---
[3] https://github.com/collab-uniba/EMTk

[4] https://stackexchange.com/sites# Last accessed February 2018.

## 6 CONCLUSIONS

Developers' communication traces, such as comments in technical Q&A sites and issue tracking systems, represent an invaluable wealth of data, ready to be mined for training predictive models on effective collaboration and communication patterns in software development. Among the information that can be extracted from such unstructured data sources, emotions and opinions are currently attracting increasing attention as they can be leveraged in empirical software engineering studies aimed at enhancing developers' productivity and well-being, informing software maintenance and evolution, and supporting effective community management. In this scenario, the software analytics community has recently investigated the potential of sentiment analysis and emotion mining as a new tool for empirical research in software development. In this paper, we contribute to this recent research trend.

## ACKNOWLEDGMENTS
This work is partially funded by the project 'EmoQuest - Investigating the Role of Emotions in Online Question & Answer Sites', funded by MIUR (Ministero dell'Università e della Ricerca) under the program "Scientific Independence of young Researchers" (SIR).